# Correlations between electrical and mechanical signals during granular stick-slip events

Karen E. Daniels[1], Caroline Bauer[2] & Troy Shinbrot[3]

[1] Dept. of Physics, North Carolina State University, Raleigh, NC, USA
[2] Institut für Experimentelle Physik, Otto-von-Guericke-Universität Magdeburg, Germany
[3] Dept. of Biomedical Engineering, Rutgers University, Piscataway, NJ, USA

Abstract:

Powders and grains exhibit unpredictable jamming-to-flow transitions that manifest themselves on geophysical scales in catastrophic slip events such as landslides and earthquakes, and on laboratory/industrial scales in profound processing difficulties. Over the past few years, insight into these transitions has been provided by new evidence that slip events may accompanied, or even preceded, by electrical effects. In the present work, we quantify the correlation between slip and the separation of electrical charges, using an archetypal granular material: photoelastic polymers. We measure a strong correlation between material displacement, acoustic emissions, and voltage. We find that the generation of voltage is associated with surface, rather than bulk properties of the granular materials. While voltage precursors are only occasionally observed in this system, there is some asymmetry in the cross-correlation between the slip and voltage signals that indicates differences between the pre-slip and post-slip dynamics.

Introduction:

It has long been reported that electrical signals are produced by material failure. As early as 1700, Bernoulli discovered that stick-slip events of liquid mercury on glass emit light[1]. Much later, in the 1930s, it was reported that adhesive tape generates light when it releases from a surface, and this was later found to extend into the x-ray spectrum[2]. Since the 1980s solid crystals[3], glasses[4], and rocks[5] have been found to produce "fractoluminescent" flashes of light during crack formation. And last year it was reported that slip events in cohesive powders also produce electrical signals[6]. Remarkably, these signals appear significantly in advance of slip events, raising the possibility that predictions of granular failures could conceivably be made.

A significant, though still inconclusive, body of research has developed into mechanisms for light emission during failure of liquids, adhesives and solids, however there is currently no theory for the mechanism underlying voltage generation in powders and grains. The granular materials studied so far are not known to be piezoelectric, the effects persist in the presence of active static elimination, and stresses are several orders of magnitude too small to produce chemical changes that have been reported elsewhere[7,8] to lead to measurable voltages.

To investigate this intriguing effect, we turn here to a granular system pioneered by RP Behringer[9] that has become emblematic of granular research: photoelastic grains. Because these materials rotate polarized light in response to mechanical stress, they permit the forces in the interior of the granular material to be visualized using crossed polarizers, and so reveal microstructural changes that occur during jamming and stick-slip transitions. Additionally, as will be described in more detail below, the grains can be directly monitored for voltage simultaneously with mechanical, acoustic, and photometric measurements to directly compare and correlate these different aspects of the dynamics.

Our experiment, shown in Fig. 1 and described in more detail in Ref. [10], is designed to mimic the geometry of a tectonic fault, and consists of a monolayer of 3 mm thick photoelastic circles and ellipses cut from PhotoStress Plus PS-3 polymer (Vishay Precision, modulus 0.21 GPa), randomly mixed to prevent crystalline packing. The circles have diameter 5.6 mm and the ellipses have major and minor axes 6.8 mm and 4.7 mm respectively; for the 10230 particles used in the experiment, this corresponds to a packing fraction of 0.79.

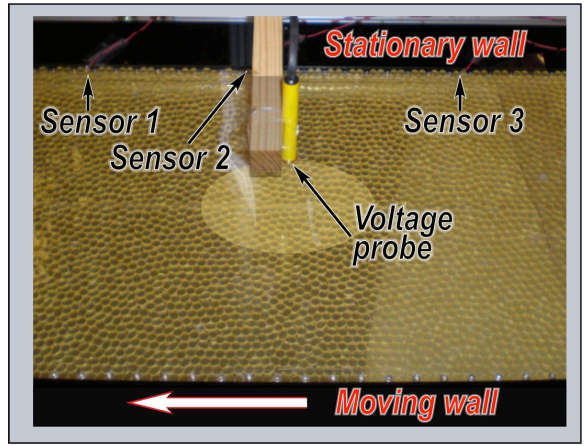

*Figure 1 – Overview of the central region of experiment (approximately one-quarter of the length, and the full width). A monolayer of mixed photoelastic disks and ellipses are sheared between two acrylic walls (black), each attached to an acrylic base plate that supports the particles from below. The shearing force and displacement are measured by instrumentation (not shown) attached to the moving wall, and the acoustic emissions from particle-slip events are monitored using three piezoelectric sensors indicated. A voltage probe is positioned 4 cm above the midline between the two plates; the approximate region facing the probe is highlighted.*

In each of 15 experimental trials, the stationary wall shown in Fig. 1 is held fixed, while the moving wall is pulled with a constant average velocity (0.3 mm/s) by a stepper motor connected to the wall by a spring. Each wall is decorated with protrusions (4 mm wide and 1 cm apart) to engage the particles at a no-slip boundary. This arrangement permits the stress on the granular bed to grow steadily until a point of Coulomb failure is reached, at which point slip occurs. The shear cell is nominally 125 cm × 26 cm in size, and the particles are weakly confined by either a 0.5 mm clear flexible PVC or 1.4 mm thick acrylic sheet in order to suppress out-of-plane buckling. The sheet is wiped with a damp cloth before the experiment to minimize triboelectrification. The chosen packing fraction represents a compromise between being dense enough for the system to exhibit stick-slip behavior, but not so stiff that buckling events dominate the dynamics given the thinness of the confining sheet (necessarily thin to permit voltage measurements).

To monitor slip, the apparatus is outfitted with sensors that allow us to continuously measure four quantities: (1) the displacement $x(t)$ of moving wall (Celesco string potentiometer SP1-12); (2) the pulling force $F(t)$ acting on the wall (Chatillon force probe DFS-025-E91-089); (3) acoustic emissions $A_1(t)$, $A_2(t)$, $A_3(t)$ from the three piezoelectric sensors shown in Fig. 1 (MSF-003-NI sensors from Piezo Systems, each approximately the size of a single particle); and (4) the voltage $V(t)$ at a height 4 cm above the surface of the particles (Trek electrostatic voltmeter 344 with probe 6000B-7C). We measure all 6 signals in synchrony using a National Instruments PXIe 6368 digitizer and data acquisition card NI 622x, with a sampling rate of 10 kHz, downsampled to 1kHz to reduce the noise level.

Results:

Typical signals from the four sets of sensors are shown in Fig. 2, for an experiment that produced both a small slip and a large slip event. The mechanical sensors (1) and (2) produce unambiguous signatures of slip, as shown in Fig. 2(a): the moving plate slips forward, and the pulling force correspondingly drops. The three acoustic sensors, whose signals are shown in Fig 2(c) and also in finer detail in Fig 3, provide an indirect measure of the amount of interparticle slip during an event. However, this measure is complicated by the presence of force chains in the system. Because force chains are highly heterogeneous, the magnitude recorded at a particular sensor location is sensitive to the immediate force chain environment of the sensor[11]. A sensor can experience larger or smaller stresses depending on whether or not it directly contacts a force chain, and the piezoelectric signals are correspondingly quite complex. For example, the bottom trace in Fig. 2(c) shows negligible evidence of the small slip event identified, and similarly in Fig's 3(b)&(d), the signals from each piezoelectric sensor strongly differ from one another. In addition, we observed acoustic emissions intermittently throughout our experiments, even when an event wasn't detected by any other sensor. These emissions presumably correspond to particle rearrangements that take place within the granular material (especially near a probe) but do not cause a rupture spanning the full size of the system. To mitigate these complications, three piezoelectric sensors were used rather than only one, and RMS averages of the three sensors were used as a metric for acoustic stress response (see Fig. 4(b)).

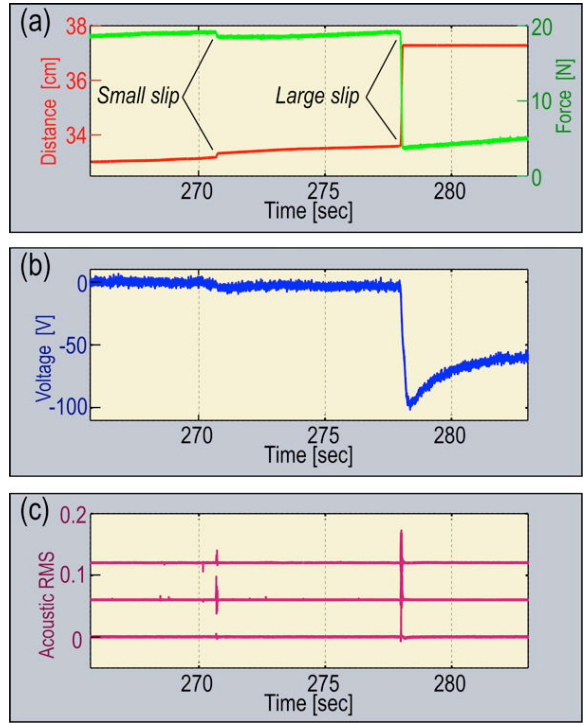

*Figure 2 – Simultaneous (a) displacement, force, (b) voltage, and (c) piezoelectric signals during a small and a large slip event. Note that while the magnitude of the voltage decrease during large slip events is easily quantifiable, it is on the order of the size of the electrical noise for small events.*

Simultaneously with these four sets of measurements, we visualize force chain rearrangements by illuminating the bed from below through one polarizing sheet, and imaging it from above through a second polarizer using a video camera. Images of the force chains immediately before and after the two events from Fig. 2, as well the changes in those chains, are shown in Fig. 3. Fig. 3 confirms the well known results that slip events are characterized both by mechanical signals (Fig's 2(a)&(c)) and force chain rearrangements (Fig. 3(a)&(c)), and that small events (Fig's 3(a)&(b)) exhibit many more rearrangements than large events (Fig's 3(c)&(d)).

Additionally, our data show that slip events produce significant and reproducible decreases in measured voltage, as displayed in Fig's 3(b)&(d). Small events involving small numbers of particle rearrangements apparently produce small voltages, comparable to electrical noise, while large events involving numerous particles correspondingly produce larger voltages. Voltage signals of this kind have not previously been reported in macroscopic grains, and photoelastic grains in particular seem to be ideal for further analysis of the root cause of the unexplained voltage generation effects, since they are both large enough to be individually tracked, and they permit stress and voltage measurements to be carried out simultaneously.

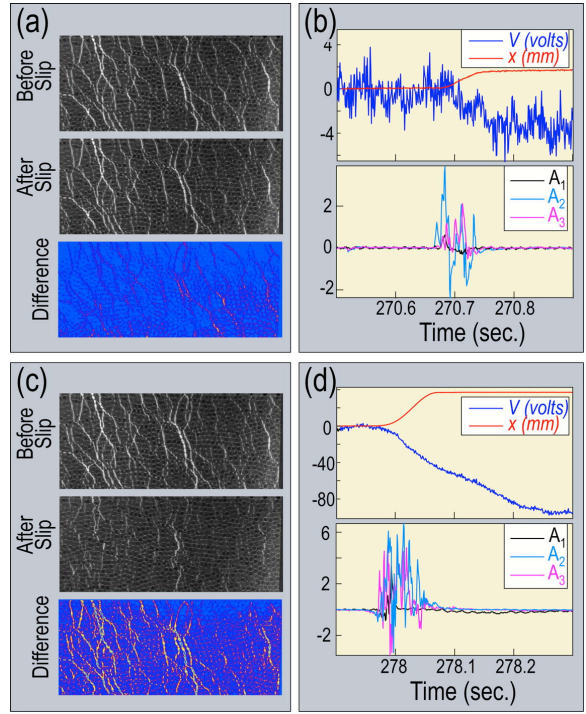

*Figure 3 – Simultaneous details of the small (a)&(b) and large (c)&(d) slip events shown in Fig. 2, seen through crossed polarizers alongside instrument signals. Sizes of events are evident in views through crossed polarizers shown in panels (a) and (c): the top two snapshot in each panel show force chains before and after a slip event, and the bottom snapshots show false colored differences between force chains before and after slip. Red indicates a large difference; blue a small difference. Voltage, distance and piezoelectric signals are all correspondingly larger for large slips than small.*

Discussion:

Armed with these diverse sources of data, we seek to examine the causes and temporal ordering of the relevant effects. In particular, we focus on three central questions. We want to determine first whether voltages are statistically correlated with slip events, second whether these voltages precede mechanical slip, and third whether the effects shown in Fig's 2 and 3 are new, or may be associated with established, e.g. piezoelectric or traditional triboelectric, causes.

For all of these questions, we need to establish a consistent criterion for a significant slip event. As we have mentioned, there are small or ambiguous events, in which a voltage signal may be confounded by noise (e.g. Fig. 3(b)). For each set of synchronized sensor measurements, we detect individual events from the temporal Fourier derivatives $x'(t)$, $F'(t)$, and $V'(t)$, in which each slip event appears as a spike. To isolate the spikes from the noise, we smooth the signal by applying a local regression over 0.1 sec, and accept all spikes above a threshold (approximately 2% of the RMS fluctuations) as events. For any event detected in at least one of the three signals, we measure the magnitudes $\Delta x$, $\Delta F$, and $\Delta V$ of each found event by fitting a tanh function to the corresponding unfiltered signal. The magnitude of the acoustic emissions is measured during the event from the RMS fluctuations $\langle A^2 \rangle$. Using this criterion, we identify 24 slip events of which 2 were detected only in $V$ (false positive), and 1 was detected only in $x$ and $F$ (false negative). Below, we use this set of events to address our three central questions.

Question 1: are voltages correlated with mechanical slip events?

In Fig. 4(a), we plot the event magnitudes $\Delta x$ and $\Delta V$. Similar results are seen for $\Delta F$ and $\Delta V$, but despite the fact that the displacement and force come from different sensors, their data are essentially redundant due to coupling through the driving spring described earlier (see also Ref. [10]).

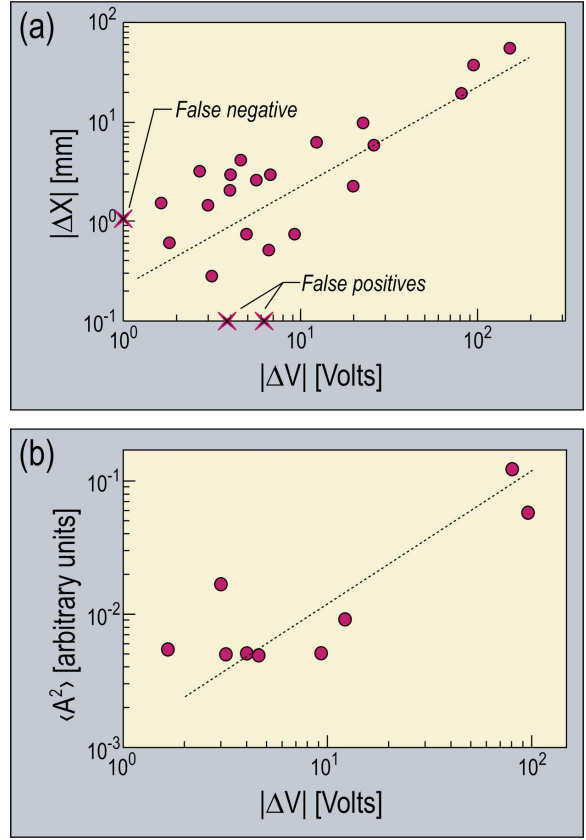

*Figure 4 – (a) Comparison between the total displacement of the moving wall, |ΔX|, and voltage drop |ΔV| for the 24 slip events seen in our experiments. The false positive and false negative events described in text are shown as red crosses placed at the corresponding magnitudes where they were detected. (b) Comparison between the acoustic RMS and the voltage drop, taken for the subset of experiments in which piezoelectric sensors were present. The dotted lines are linear dependences, for comparison.*

Fig. 4(a) confirms the impression from Fig. 3 that voltage and slip event magnitudes are strongly correlated. Indeed, larger events appear to produce larger voltages: a linear relationship is shown as a dotted line for comparison in Fig. 3; a linear least squares fit produces a correlation coefficient $r = 0.82$. Statistically, this apparent correlation can be further quantified using a Fisher exact test, which provides a measure of likelihood that a correlation could be due to random chance. The most conservative possible Fisher exact test indicates that voltage and mechanical slip are very strongly correlated, with a probability of $10^{-8}\%$ that the observed correlation is random. By 'most conservative', we mean the largest conceivable estimation of randomness as a cause. We obtain this conservative test by assuming that a true negative corresponds to any contiguous period of time before a voltage that lacks a slip event. If we were instead to define a less conservative measure, say, 1 second intervals during which a voltage was not detected and a slip did not occur, we would obtain very large numbers of true negatives (in the hundreds to thousands) that would correspondingly produce infinitesimally small ($10^{-24}$ or lower) Fisher exact likelihoods that the result could be random. For any analysis of the data, the voltage dips measured are very strongly correlated with mechanical slip, and the likelihood that this correlation could be due to random chance is exceedingly small.

Before examining the dynamics that give rise to this effect, we need to recognize that the piezoelectric sensors – which are electrical in nature – could conceivably interfere with our voltage measurements. To exclude this possibility, we performed half of the experiments shown in Fig. 4(a) with the piezoelectric sensors removed. We found no difference between the mechanical or voltage measurements with and without the sensors, or for the different covering materials. For the RMS averages of piezoelectric sensor output and voltages measured for those tests using the sensors, we also observe an approximately linear dependence, as shown as a dotted line in Fig. 4(b). Here, we obtain a correlation coefficient of $r^2 = 0.85$. By all of these measures – qualitative observation, quantitative Fisher exact results, and quantitative response curves – we conclude that voltage drops are strongly correlated with mechanical slip.

Question 2: do voltages precede mechanical slip?

From the event magnitudes, we conclude that voltage signals are strongly correlated with slip events in our experiment. Our second question is whether the signals precede mechanical slip, and here the results are more equivocal. We did observe some events in which the voltage changes preceded the mechanical response, as shown in Fig. 5(a). For this one experiment, the packing fraction was slightly less dense (10180 instead of 10230 particles), and the event was accompanied by out-of-plane buckling, with particles leaving the bed. The event is additionally peculiar in that the voltage ends up higher, rather than lower, at the end of the event. Nonetheless, it presents the possibility of precursor voltages, and is similar to plots of rates of liquid bridge extinction that precede slip in simulations[12] as well as to growth of disorder in experiments involving avalanching grains[13].

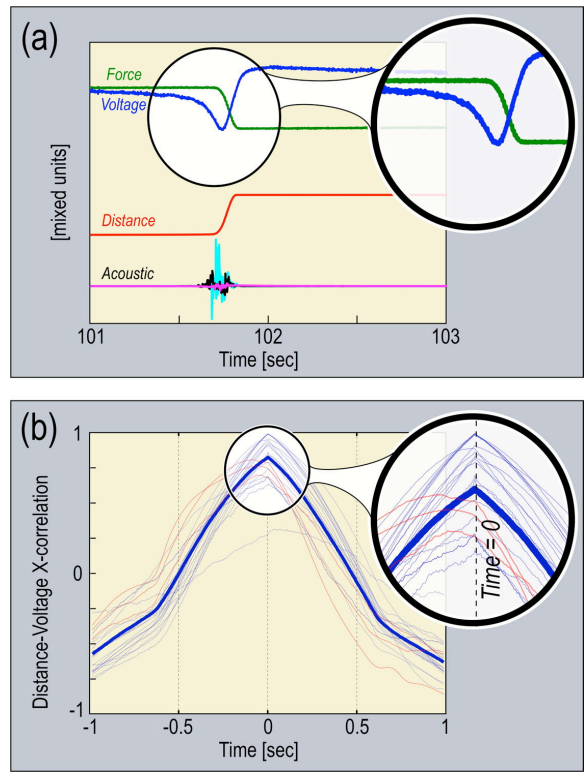

*Figure 5 – Correlations between voltage and mechanical signals. (a) Example of an event in which part of the voltage signal preceded the mechanical slip, highlighted in the enlarged inset. (b) Cross correlations between position x(t) and voltage V(t). Cross correlations for each individual event are plotted as thin lines, and the average over all 24 events is plotted as a thick line. Cases in which skew is seen and the peak appears to indicate a voltage preceding the slip are shown as light red lines.*

On the other hand, we stress that putative precursors such as the one shown in Fig. 5(a) are not the norm – much more common are cases such as the slip events of Fig. 2 and 3, where no precursor is apparent. To evaluate the predictive strength of putative precursors, we have therefore performed cross-correlations of all of the slip events detected in our experiments. As shown in Fig. 5(b), this analysis shows skew in some of the individual cross-correlation ⟨x(t) V(t+Δt)⟩ plots, identified in red in the figure, indicating that there may be differences between pre-slip and after-slip voltage dynamics. No consistent correlation between skew and event magnitude was observed. Ultimately, we conclude that although there remain unexplained Physics underlying the voltage changes seen, there is no significant peak that could indicate a consistent voltage precursor that would provide predictive power, at least on the scale of our experiment.

Question 3: are the voltages caused by a new effect?

The cause of the voltages detected remain unclear: they could be related to triboelectrification, piezoelectrification, cohesive or ordering effects as described in prior work[13], or some new and unexplained effect. While we have not been able to isolate the underlying cause in this work, we have performed a final set of simple tests to evaluate the properties of individual photoelastic disks in contact, which we hope will lay the groundwork for future investigations.

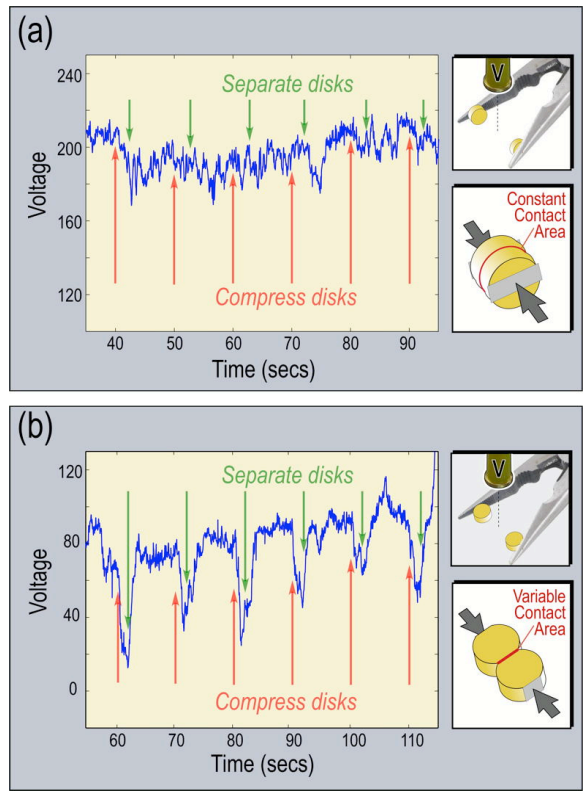

*Figure 6 – Voltage measured near pairs of compressed photoelastic disks in two orientations. (a) The disks are contacted with flat sides facing one another by gluing one bead to each jaw of a pliers as shown in the upper right inset. The beads are repeatedly compressed to about half their rest thickness, at a distance of 1 cm from the voltage probe ("V" in inset), but no consistent voltage signal is observed. Since surface contact area doesn't change with compression in this orientation, this indicates that bulk stresses do not generate voltages. (b) The disks are compressed exactly as before, but with curved sides facing, as shown in the upper right inset. Here the contact area varies with compression, and a reproducible voltage signal is seen. This indicates that surface stresses are associated with voltages, but bulk stresses are not.*

As shown in Fig. 6, we compress two disks within the jaws of needle-nosed pliers held 1 cm in front of the voltage probe. We perform these tests with disks glued to the jaws in two orientations: with flat-to-flat contacts as shown in the inset to Fig. 6(a), and with curve-to-curve contacts as shown in the inset to Fig. 6(b). The glue permits us to separate as well as compress the particles, and these orientations are used because in the first case, the disks can be compressed without increasing their contact area, while in the second case, as the disks are compressed, their contact area will increase (see lower right insets in each case).

In both cases, every 10 seconds the disks are repeatedly compressed to about half their rest thickness and then separated. The compression and separation are performed manually, last about a second and are performed with care to maintain near constant separation from the voltage probe. In all tests, the voltage drifts upward in magnitude over the first half minute for an unknown reason; the data in Fig. 6 are shown after the voltage has stabilized. Fig. 6(a) shows typical data with flat, fixed area, contacts, and displays no detectable response to either compression or separation. This lack of response provides evidence that the voltage detected in our earlier experiments is not associated with a bulk, e.g. piezoelectric, effect.

Fig. 6(b) shows distinctly different behavior: every time the disks are compressed, the voltage drops by about 50V, and every time the disks are separated, the voltage grows by about the same amount. Some noise – as well as the unexplained drift mentioned earlier – remains, but the voltage generation associated with the second orientation seems unambiguous, and was reproduced in repeated trials. The source of the voltage signal is, as we have mentioned, unknown. However, since it is not seen in Fig. 6(a), the voltage seems to be associated with surface, rather than bulk, phenomena.

Both arrangements shown in Fig. 6 are free of confining surfaces, so the effect does not seem to be caused by rubbing against top or bottom plates of the shearing cell shown in Fig. 1. Moreover, in the shearing experiments, as we mentioned, we wiped the top surface with a damp cloth, used two different covering plate materials, and we additionally monitored the relative humidity present in the laboratory and observed indistinguishable voltage-slip correlations on four days at humidities in the range 20-40% RH.

It is possible that the increased contact with the plier surface in Fig. 5(b) could generate a voltage, however two facts bear mention. First, the plier itself was insulated from ground, so there is no source of net voltage, hence even if a potential difference were generated between plier and disk, the voltage measured at the probe should remain constant. Second, known contact electrification mechanisms generate measurable voltage differences *when surfaces are separated* from one another but not when they are brought into contact. By contrast, in our tests making and breaking contact produced voltages that were mirror images of one another. We therefore conclude that whatever effect drives the voltage production seen here, it seems to be caused by surface phenomena that are difficult to explain using existing piezoelectric or triboelectric models.

Conclusion:

To conclude, we have performed shearing and compression experiments using photoelastic polymers, and we find that slip events are strongly correlated with voltage signals. The root cause of these signals remains unclear, but it appears that the signals are not associated with piezoelectric or triboelectric phenomena as they are currently understood. Rather, the voltages seem to be produced through changes in stresses associated with surfaces in contact. Prior work indicates that slip may be preceded by the making and breaking of contacts in disordered beds, and there may be opportunity for advancement of understanding of this new voltage generation phenomenon at the intersection between the fields of soft matter physics and traditional materials science.

Acknowledgments:

The authors thank Theo Siu for his invaluable assistance, and are grateful to the National Science Foundation (grant DMR-1206808), the Alexander von Humboldt-Stiftung, and the DAAD-RISE program for financial support.